\input harvmac
\input amssym

\def\coeff#1#2{\relax{\textstyle {#1 \over #2}}\displaystyle}

\def\oneone{\rlap 1\mkern4mu{\rm l}}

 \def\cK{{\cal K}}
\def\cL{{\cal L}} 
\def\cN{{\cal N}} 
 
\def\cR{{\cal R}} 
 
\def\cW{{\cal W}}

\def\bfone{\relax{\rm 1\kern-.35em 1}}
\def\IC{\relax\,\hbox{$\inbar\kern-.3em{\rm C}$}}
\def\ID{\relax{\rm I\kern-.18em D}}
\def\IF{\relax{\rm I\kern-.18em F}}
\def\IH{\relax{\rm I\kern-.18em H}}
\def\II{\relax{\rm I\kern-.17em I}}
\def\IN{\relax{\rm I\kern-.18em N}}
\def\IP{\relax{\rm I\kern-.18em P}}
\def\IQ{\relax\,\hbox{$\inbar\kern-.3em{\rm Q}$}}

\def\IR{\relax{\rm I\kern-.18em R}}
\font\cmss=cmss10 \font\cmsss=cmss10 at 7pt
\def\ZZ{\relax\ifmmode\mathchoice
{\hbox{\cmss Z\kern-.4em Z}}{\hbox{\cmss Z\kern-.4em Z}}
{\lower.9pt\hbox{\cmsss Z\kern-.4em Z}}
{\lower1.2pt\hbox{\cmsss Z\kern-.4em Z}}\else{\cmss Z\kern-.4em
Z}\fi}
\def\Tr{{\rm Tr}\, }
\def\ZZ{\Bbb{Z}} 
\def\IC{\Bbb{C}}
\def\ID{\Bbb{D}}
\def\IF{\Bbb{F}}
\def\IH{\Bbb{H}}
\def\II{\Bbb{I}}
\def\IN{\Bbb{N}}
\def\IP{\Bbb{P}}
\def\IQ{\Bbb{Q}}
\def\IR{\Bbb{R}}
%
%
 %
\lref\CandelasJS{
P.~Candelas and X.~C.~de la Ossa,
``Comments On Conifolds,''
Nucl.\ Phys.\ B {\bf 342}, 246 (1990).
}
%
\lref\RomansAN{
L.~J.~Romans,
``New Compactifications Of Chiral N=2 D = 10 Supergravity,''
Phys.\ Lett.\ B {\bf 153}, 392 (1985).
}
%
\lref\KlebanovHH{
I.~R.~Klebanov and E.~Witten,
``Superconformal field theory on threebranes at a Calabi-Yau  singularity,''
Nucl.\ Phys.\ B {\bf 536}, 199 (1998)
[arXiv:hep-th/9807080].
}
%
\lref\PilchEJ{
K.~Pilch and N.~P.~Warner,
``A new supersymmetric compactification of chiral IIB supergravity,''
Phys.\ Lett.\ B {\bf 487}, 22 (2000)
[arXiv:hep-th/0002192].
}
%
\lref\PilchFU{
K.~Pilch and N.~P.~Warner,
``N = 1 supersymmetric renormalization group flows from IIB supergravity,''
Adv.\ Theor.\ Math.\ Phys.\  {\bf 4}, 627 (2002)
[arXiv:hep-th/0006066].
}
\lref\PilchYG{
K.~Pilch and N.~P.~Warner,
``N = 1 supersymmetric solutions of IIB supergravity from Killing spinors,''
arXiv:hep-th/0403005.
}
%
\lref\GowdigereJF{
C.~N.~Gowdigere, D.~Nemeschansky and N.~P.~Warner,
``Supersymmetric solutions with fluxes from algebraic Killing spinors,''
arXiv:hep-th/0306097.
}
%
\lref\PilchJG{
K.~Pilch and N.~P.~Warner,
``Generalizing the N = 2 supersymmetric RG flow solution of IIB
supergravity,''
Nucl.\ Phys.\ B {\bf 675}, 99 (2003)
[arXiv:hep-th/0306098].
}
%
\lref\NemeschanskyYH{
D.~Nemeschansky and N.~P.~Warner,
``A family of M-theory flows with four supersymmetries,''
arXiv:hep-th/0403006.
}
%
\lref\tianyaua{
G.~Tian and S.~T.~Yau,
``Complete K\"ahler Manifolds with Zero Ricci Curvature. I,''
Journal of the American Mathematical Society, vol. 3, number 3, July 1990, p. 579
}
%
\lref\tianyaub{
G.~Tian and S.~T.~Yau,
``Complete K\"ahler Manifolds with Zero Ricci Curvature. II,''
Invent. Math. 106, 27-60 (1991)
}
%
\lref\DouglasZJ{
M.~R.~Douglas and B.~R.~Greene,
``Metrics on D-brane orbifolds,''
Adv.\ Theor.\ Math.\ Phys.\  {\bf 1}, 184 (1998)
[arXiv:hep-th/9707214].
}
%
\lref\DouglasSW{
M.~R.~Douglas and G.~W.~Moore,
``D-branes, Quivers, and ALE Instantons,''
arXiv:hep-th/9603167.
}
%
\lref\CorradoWX{
R.~Corrado, M.~Gunaydin, N.~P.~Warner and M.~Zagermann,
``Orbifolds and flows from gauged supergravity,''
Phys.\ Rev.\ D {\bf 65}, 125024 (2002)
[arXiv:hep-th/0203057].
}
%
\lref\GranaXN{
M.~Grana and J.~Polchinski,
``Gauge / gravity duals with holomorphic dilaton,''
Phys.\ Rev.\ D {\bf 65}, 126005 (2002)
[arXiv:hep-th/0106014].
}
%
\lref\BeckerGJ{
K.~Becker and M.~Becker,
``M-Theory on Eight-Manifolds,''
Nucl.\ Phys.\ B {\bf 477}, 155 (1996)
[arXiv:hep-th/9605053].
}
%
\lref\BerkoozKM{
M.~Berkooz, M.~R.~Douglas and R.~G.~Leigh,
``Branes intersecting at angles,''
Nucl.\ Phys.\ B {\bf 480}, 265 (1996)
[arXiv:hep-th/9606139].
}
%
\lref\BeckerKB{
K.~Becker, M.~Becker and A.~Strominger,
``Five-branes, membranes and nonperturbative string theory,''
Nucl.\ Phys.\ B {\bf 456}, 130 (1995)
[arXiv:hep-th/9507158].
}
%
\lref\LeighEP{
R.~G.~Leigh and M.~J.~Strassler,
``Exactly marginal operators and duality in four-dimensional N=1 supersymmetric
gauge theory,''
Nucl.\ Phys.\ B {\bf 447}, 95 (1995)
[arXiv:hep-th/9503121].
}
\lref\JohnsonZE{
C.~V.~Johnson, K.~J.~Lovis and D.~C.~Page,
``The K\"ahler structure of supersymmetric holographic RG flows,''
JHEP {\bf 0110}, 014 (2001)
[arXiv:hep-th/0107261].
}
%
\lref\FreedmanGP{
D.~Z.~Freedman, S.~S.~Gubser, K.~Pilch and N.~P.~Warner,
``Renormalization group flows from holography supersymmetry and a  c-theorem,''
Adv.\ Theor.\ Math.\ Phys.\  {\bf 3}, 363 (1999)
[arXiv:hep-th/9904017].
}
%
\lref\CorradoBZ{
R.~Corrado and N.~Halmagyi,
``N = 1 field theories and fluxes in IIB string theory,''
arXiv:hep-th/0401141.
}
\lref\GubserIA{
S.~Gubser, N.~Nekrasov and S.~Shatashvili,
``Generalized conifolds and four dimensional N = 1 superconformal  theories,''
JHEP {\bf 9905}, 003 (1999)
[arXiv:hep-th/9811230].
}
%
\lref\KhavaevFB{
A.~Khavaev, K.~Pilch and N.~P.~Warner,
``New vacua of gauged N = 8 supergravity in five dimensions,''
Phys.\ Lett.\ B {\bf 487}, 14 (2000)
[arXiv:hep-th/9812035].
}
\lref\GNPW{C.~Gowdigere, D.~Nemeschansky, K.~Pilch, and N.P.~Warner, to appear.}
%



\Title{
\vbox{
\hbox{\tt hep-th/0406147}
}}
{\vbox{\vskip -1.0cm
\centerline{\hbox{The Complex Geometry of Holographic Flows}}
\vskip 8 pt
\centerline{\hbox{of Quiver Gauge Theories }}}}
\vskip -.3cm
\centerline{Nick Halmagyi${}^{(1)}$,  Krzysztof Pilch${}^{(2),(1)}$%
\footnote{${}^\dagger$}{On sabbatical leave.}, }
\medskip
\centerline{Christian R\"omelsberger${}^{(1)}$ and
Nicholas P.\ Warner${}^{(1)}$}
\bigskip
\centerline{{${}^{(1)}$\it Department of Physics and Astronomy}}
\centerline{{\it University of Southern California}}
\centerline{{\it Los Angeles, CA 90089-0484, USA}}
\medskip
\centerline{{${}^{(2)}$\it Departament ECM, Facultat F\'\i sica }}
\centerline{{\it Universitat de Barcelona}}
\centerline{{\it Diagonal 647}}
\centerline{{\it 08028 Barcelona, Spain}}
\medskip

\bigskip
\bigskip
We argue that the complete Klebanov-Witten flow solution must be
described by a Calabi-Yau metric on the conifold, interpolating between
the orbifold at infinity and the cone over $T^{(1,1)}$ in the interior.
We show that the complete flow  solution is characterized completely by a single,
simple, quasi-linear, second order PDE, or ``master equation,'' in two variables.  
We show that the Pilch-Warner flow solution is almost Calabi-Yau: It has a
complex structure, a hermitian metric, and a holomorphic $(3,0)$-form that is a 
square root of the volume form.  It is, however, not K\"ahler.  We discuss
the relationship between the master equation derived here
for Calabi-Yau geometries and such equations encountered elsewhere
and that govern supersymmetric backgrounds with multiple, independent fluxes. 

\vskip .3in
\Date{\sl {June 2004}}

\vfill\eject

\newsec{Introduction}

One of the first families of physical, supersymmetric holographic flows
to be studied were those   \refs{\KlebanovHH,\GubserIA} describing
the massive, $\cN=1$ supersymmetric flows from $\cN=2$ 
quiver gauge theories.  These flows were shown to go to Leigh-Strassler
type fixed point theories in the infra-red, and the holographic duals
of these fixed points were also identified.   It has remained one of
the unsolved problems in holography to construct these solutions explicitly,
and while this is hard for the general quiver theories, one might expect
to be able to solve this problem for the $\widehat A_1$ quiver
of \KlebanovHH\ because of its much higher level of symmetry.
Despite numerous attempts, there has been rather little progress 
in finding the flow solution within supergravity.  

Shortly after the work of  \refs{\KlebanovHH,\GubserIA}, new supergravity
fixed points were discovered in \KhavaevFB, and the corresponding flow solutions
were obtained in \FreedmanGP.   While these flows were originally 
investigated for $\cN=4$ Yang-Mills theory, it was noted at the time
that these solutions were also flows in the untwisted sector of the
$\cN=2$  superconformal quiver gauge theories obtained via orbifolds.
These supergravity solutions were originally constructed in five dimensions,
but their complete ten-dimensional analogs were subsequently constructed 
in \refs{\PilchEJ, \PilchFU}, and then significantly generalized in a recent paper
\PilchYG.    While these supergravity solutions were explicitly known, the problem
was that their underlying geometric structure was far from evident:  The solutions
were obtained via the magic of consistent truncation, and not through a 
clearer geometric principle.  The complication is that these solutions involve 
multiple, non-trivial fluxes and are not some straightforward decoration
of a Calabi-Yau geometry.

Finally, in \CorradoWX\ it was argued that the  holographic flow solutions 
of \refs{\KlebanovHH,\GubserIA} and \refs{ \KhavaevFB, \FreedmanGP}
should be related via a continuous family of duality transformations.
Further evidence for this was found in the field theory analysis of
\CorradoBZ.

Our purpose in this paper is to re-examine and  understand the geometry
of both of these families of flows using, in particular, the ideas of 
\refs{\GowdigereJF,\PilchJG,\PilchYG,\NemeschanskyYH} for
classifying the ``supersymmetry bundles.'' 
For the Klebanov-Witten flow we will argue that the internal geometry of
the entire flow solution must live on the singular conifold.  That is, the
internal manifold must be governed by a   Ricci-flat, K\"ahler metric 
on the conifold, and that this metric must interpolate between the orbifold,
$\IC^2/\ZZ_2 \times \IC$ at infinity and the cone over $T^{(1,1)}$ in the
interior.  The orbifold singularity at infinity  is thus blown up as one moves
radially inwards.  We then proceed to characterize this geometry exactly in terms
of a ``master equation'' that is a single, second-order, quasi-linear PDE in 
two variables.  This characterization of the solution represents a significant simplification
over the more traditional Monge-Amp\`ere equation (which is strongly non-linear,
and sixth-order).

For the Pilch-Warner (PW) flow we re-examine the geometry underlying the
general class of solutions in \PilchYG.  We will show that the internal manifold
has an {\it integrable} complex structure and a holomorphic $(3,0)$-from
such that $\Omega \wedge \overline \Omega$ is the volume form of the manifold.
The internal geometry is thus {\it almost} Calabi-Yau, except that it
is {\it not} K\"ahler:  The K\"ahler form is not closed.   It was also
shown in \PilchYG\ that this non-trivial flux geometry was governed by
a  single, second-order, quasi-linear PDE,  or ``master equation.''  Indeed,
this master equation is very simply related to the one that we find here for the
Calabi-Yau geometry of the Klebanov-Witten flow.  

This observation now provides new insight into the broader class
of results contained in \refs{\GowdigereJF,\PilchJG,\PilchYG,\NemeschanskyYH}.
These were all supersymmetric, multiple flux solutions, and were all governed
by a single master equation similar to those encountered here.  Our work
here suggests that all these multiple flux solutions are indeed some relatively
simple deformation of a Calabi-Yau condition \GNPW.

In section 2 we will use a combination of field theory, supergravity and brane-probe
analysis to argue that the complete Klebanov-Witten flow should be described 
by a Calabi-Yau metric on the singular conifold. In section 3 we look at the
conifold geometry in some detail, and use the symmetries of the flow
that we seek to pin down the general form of the metric on the internal space.
Having obtained a suitably general metric Ansatz, in section 4 we use
the techniques of  \refs{\GowdigereJF,\PilchJG,\PilchYG,\NemeschanskyYH} 
to solve the Killing spinor equations, and thereby obtain the Calabi-Yau metric
in terms of the solution of the master equation.  In section 5 we
repeat the analysis, but this time more systematically in terms of the
special holonomy of the spin connection.   The end-result  is the same as
that of section 4, but one sees very explicitly how each equation emerges 
from distinct conditions on the holonomy.  Section 5 also contains
a more detailed discussion of the complex structure and coordinates, and
contrasts our approach with adaptions of the ideas of Tian and Yau, which
would lead to the Monge-Amp\`ere equation.
Section 6 contains our geometric re-examination of the geometry of the PW flow,
and section 7 contains some final remarks.

\newsec{Field Theory}

The UV point of the Klebanov-Witten flow is the $\cN =2$,  ${\widehat A}_{1}$  quiver 
gauge theory. This has gauge group $SU(N)\times SU(N)$ and
also has two bi-fundamental hypermultiplets. In terms of $\cN =1$ multiplets,
the two $\cN=2$ vector multiplets contain
 two $\cN=1$ adjoint chiral superfields $\Phi_1$, $\Phi_2$ while each of the hypermulitplets yields a pair of chiral multiplets $(A_1,B_1)$ and $(B_2,A_2)$.
The chiral superfields $A_1$, $B_2$ lie in the $(N,\bar N)$ representation
while $B_1$, $A_2$ are in the $(\bar N,N)$. 
The superpotential is
\eqn\ntwosuperpot{\eqalign{W&=\Tr\left[ \left(\matrix{\Phi_1 & 0 \cr 0 & \Phi_2}\right) \
\left[\left(\matrix{0 & A_1 \cr A_2 & 0}\right),\left(\matrix{0 & B_2 \cr B_1 & 0}\right)\right] 
\right] \cr &=\Tr\left[ \Phi_1(A_1B_1-B_2A_2)\right]  +
\Tr\left[ \Phi_2(A_2B_2-B_1A_1)\right] \,.}}
This theory has a $SU(2)\times U(1)$ $\cR$-symmetry together with a global 
$SU(2)$ symmetry under which the two hypermultiplets form a doublet.

The Klebanov-Witten flow is driven by the relevant operator
\eqn\massdef{\Delta W~=~ \coeff{1}{2}\, m\,  \Tr[\Phi_1^2] ~-~ 
\coeff{1}{2}\, m\,   \Tr[\Phi_2^2]}
which breaks the $\cN=2$ supersymmetry of the
vector multiplet. The IR point of this flow has only the bi-fundamental fields,
the adjoint fields having been integrated out. It has a superpotential
\eqn\conisup{W_{\cN=1}~=~  \lambda \, \Tr(A_1B_1B_2A_2-
A_1A_2B_2B_1)\,.}
The methods of Leigh and Strassler  \LeighEP\ suggest that
this is a non-trivial CFT  \KlebanovHH. 

This theory still has the $SU(2)$ global symmetry
but the $SU(2) \times U(1)$ $\cR$-symmetry is broken to a $U(1)$ $\cR$-symmetry. 
There are also two $\ZZ_2$ $\cR$-symmetries. One of them is the symmetry 
exchanging the  two gauge groups, while the other symmetry is charge conjugation 
\KlebanovHH.   The obvious interchange symmetry  is:
\eqn\obvintersymm{ A_1  \leftrightarrow A_2 \,, \qquad B_1  \leftrightarrow B_2\,,
\qquad \Phi_1  \leftrightarrow   \Phi_2 \,.}
However, this is not a symmetry of the perturbation \massdef.   To get a
symmetry of the complete perturbed superpotential we need to combine the
interchange with an $\cR$-symmetry transformation:
\eqn\intersymm{ A_1  \to i\, A_2 \,, \quad A_2  \to i\, A_1\,, \quad 
B_1  \to i\, B_2 \,, \quad B_2  \to i\, B_1\,, \quad 
\Phi_1  \leftrightarrow   \Phi_2  \,,}
under which $W + \Delta W \to -(W + \Delta W)$, and hence the action is invariant.
We will need this particular symmetry later.

\subsec{The brane-probe theory}

A fundamental feature of D-brane physics is that since they are BPS objects, 
D-branes are free to move in their tranverse directions
without feeling a force from parallel branes. This  manifests itself as a moduli 
space  in the gauge theory living on the branes. In this section we use this
Higgs moduli space along the Klebanov-Witten flow to obtain 
important insights into the geometry of the holographic dual theory.

When a single brane moves off the stack and probes the internal geometry, 
the gauge group is broken according to:

\eqn\gaugebraking{SU(N+1)\times SU(N+1) \rightarrow SU(N)\times U(1)
\times SU(N)\times U(1)\,.}
Since $N \gg 1$, the $SU(N)$ factors produce the 
background whereas the $U(1)$ factors correspond to the brane-probe theory,
and their gravitational effects can be neglected. 
If the probe brane is located far away from the rest of the branes, the W bosons are 
very heavy and the $U(1)$ part of the theory decouples. We now concentrate on 
the decoupled $U(1)\times U(1)$ gauge 
theory. The diagonal $U(1)$ factor is just a free $U(1)$ and we will drop it.

The anti diagonal $U(1)$ couples to four scalars $A_1$, $A_2$, $B_1$ and $B_2$.
The fields $A_1$ and $B_2$ have a $U(1)$ charge $+1$, wheras $A_2$ and $B_1$ 
have  charge $-1$. 
Since these matter fields have large vevs, the remaining $U(1)$ 
can be integrated out. This corresponds to a K\"ahler quotient. 
Alternatively, the K\"ahler quotient is equivalent to solving
D-flatness modulo this $U(1)$.

The resulting space of vacua is parameterized by the gauge invariant variables
\eqn\gaugeinvvar{z_1=A_1B_1\,,\quad z_2=B_1B_2\,,\quad z_3=A_2A_1\,
\quad{\rm and}\quad z_4=B_2A_2\,.}
These variables satisfy 
\eqn\coni{z_1 z_4-z_2 z_3=0,}
the defining equation of the singular conifold.  

This result was deduced from  the structure of the $F$-terms of the field theory, and 
one should note that, provided that there are no non-perturbative corrections,
this analysis is valid not only valid for the IR fixed point, but
for all points along the RG flow.    As a result, the moduli space of the brane probes
for the complete flow should be described by the conifold, and in particular
it should inherit its complex structure.   Since the $D$-terms of the field
theory action undergo strong renormalization, the metric of the field theory, and
of the brane probes will be non-trivial.     We therefore infer that as far as 
the brane probes are concerned, the flow will be described by
some hermitian metric on the conifold.

To get a better picture of what is happening in the UV, it  
is convenient to change coordinates to:
\eqn\wfromz{w_1=z_1+z_4\,,\quad w_2=i(z_2+z_3)\,,\quad w_3=z_2-z_3\,
\quad{\rm and}\quad w_4=i(z_1-z_4)\,, }
which satisfy:
\eqn\coniw{w_1^2~+~ w_2^2~+~w_3^2~+~w_4^2~=~0\,.}
Solving the F-flatness condition in \ntwosuperpot\  and \massdef\ for the $\Phi_i$ yields 
\eqn\fflatone{m\, \Phi_1~=~ -( A_1B_1-B_2A_2),\qquad  
m\, \Phi_2  ~=~ A_2B_2-B_1A_1}
and since the probe theory is Abelian, we can write this as 
\eqn\Phiw{w_4 = im\,  \Phi_1=  im\, \Phi_2.}
Thus the moduli space has a preferred coordinate, $w_4$, that is dual to the
vev of the scalar fields in $\cN=2$ vector multiplet.  Note that at the UV
fixed point, when $m =0$, one has $w_4 =0$ and $\Phi_1 + \Phi_2$ is
a freely choosable, independent field.  This means that in the
UV the probe moduli space is simply the orbifold, $\IC^2/\ZZ_2 \times \IC$,
and that the conifold has degenerated.  
At the UV point, the scalar kinetic term and the brane probe metric have the
trivial Euclidean, flat form while in the IR the brane-probe moduli 
space lies on the singular conifold
with the Calabi-Yau metric obtained from the cone over $T^{(1,1)}$ \KlebanovHH.
This naturally leads one to consider the family of manifolds
\eqn\conifoldwithepsilon{w_1^2~+~ w_2^2~+~w_3^2~+~\epsilon^2\, w_4^2~=~0\,. }
For an arbitrary, non-zero value of $\epsilon$ we can rescale $w_4$ and reproduce \coniw, 
however the K\"ahler metric on the moduli space will make it possible to determine 
the size of the constant $\epsilon$. Thus by varying $\epsilon$ between 
$0$ and $1$ we can  interpolate between $\IC^2/\ZZ_2 \times \IC$ and the singular 
conifold.   

Whilst the previous calculation is not  new or difficult, our interpretation of 
the result is our guide to the construction of the dual geometry of the whole flow.
Namely, it suggests that the holographic dual of the entire flow has the complex 
structure of the conifold. However the K\"ahler metric on this conifold has to
be squeezed so as to ultimately degenerate to the orbifold in the UV.
To obtain the exact result we need to determine a squeezed K\"ahler 
metric on the conifold that respects the symmetries of the flow.  One might hope
to get this metric from the field theory, however quantum corrections mean that 
the metric on the  brane-probe moduli  space doesn't necessarily coincide with 
the metric obtained from the K\"ahler quotient of the field theory metric
obtained from the UV limit \refs{\DouglasZJ}.

\subsec{Supergravity}

Having gained some insight from the field theory, we now turn more directly to
the gravity side of the duality. In addition to the dual geometry we must consider 
the possibility of non-trivial profiles for the
fluxes and the dilaton of IIB supergravity.

The $\cN=2$ quiver gauge theory can be realized 
in string theory by D3-branes probing a $\IC^2/\ZZ_2$ orbifold singularity 
\refs{\DouglasSW}. The mass deformation \massdef\ is then described by 
a twisted-sector, closed string mode.
The corresponding twisted-sector, closed string modes of type IIB theory on an ALE 
orbifold come in six-dimensional $\cN=(2,0)$ tensor multiplets associated
to the blow-up of a two-cycle.  Each
such multiplet contains  five scalars:  Three of these scalars are  
 hyper-K\"ahler deformations while the remaining two scalars describe
the NS and RR two-form flux through the two-cycle. Putting D3-branes on
the singularity breaks supersymmetry to four-dimensional $\cN=2$ supersymmetry.
The two scalars describing the two-form flux then couple at first order to the 
relative gauge couplings and theta angles of the gauge theory. On the other 
hand the three scalars describing the hyper-K\"ahler blow-up modes couple to
the D-terms and F-terms \refs{\CorradoWX}.
This shows, that at leading order our mass deformation corresponds to a 
geometric deformation \conifoldwithepsilon\ of the orbifold and that the
 two-form potentials and three-form fluxes are not being turned on. 
 Also, to leading order  the dilaton-axion is not running.  We will
therefore seek supergravity  with no such fluxes, and with a trivial
dilaton and axion.
  
The brane-probe analysis also yields more evidence for this proposal.
If the holographic dual theory had non-zero three-form flux, then one  would
expect there to be a  potential for the brane-probe that would restrict it to a 
subspace of less than six dimensions \JohnsonZE.
Since we found a six-dimensional brane-probe moduli space, it seems natural 
that there should be no three-form fluxes.

We can now use the fact that our D3-branes can probe the whole geometry.
In order for the probe brane to preserve supersymmetry, the Killing spinor,
$\epsilon$, in the holographic dual theory has to satisfy 
\refs{\BeckerKB,\BerkoozKM}:
\eqn\beckerproj{\Gamma^{1234}\epsilon= \pm i \, \epsilon}
(for one uniform choice of sign, depending upon conventions) 
at the location of the probe brane. Since the probe brane can be moved 
anywhere in the geometry, equation \beckerproj\ has to hold at all points.
We can utilise results from \GranaXN\ where it was established that 
such solutions are of Becker-type. 

This means that  the internal manifold is Calabi-Yau, 
in agreement with our conclusions from the previous section.
In Becker type compactifications the warp factor in front
of the metric  parallel to the branes is $Z^{- {1 \over 2}}$, where
$Z$  satisfies a harmonic equation, and the five-form flux is related to the 
warp factor by: 
\eqn\ffromhzero{f=-{1\over 4 \, Z}\,,}
where the five-form flux is given by
\eqn\fiveformflux{F= (1+\ast)\,df\wedge dx^1\wedge dx^2\wedge 
dx^3\wedge dx^4\,.}
There could be a holomorphic dilaton or imaginary self dual three-form fluxes, 
but we already agued that they vanish.

The fact that the internal geometry is Calabi-Yau implies the that
the Killing spinors can be determined using three independent helicity projectors
of the form:
\eqn\genhelproj{ \Pi^{(ab)}~=~  {1\over 2} \big(\oneone - i \,\Gamma^{a}\Gamma^{b} \big)\,,}
for some frame indices, $a$ and $b$, on the Calabi-Yau manifold.
These helicity projectors have the effect of isolating the spinor that
is a singlet under the $SU(3)$ holonomy. 

In summary, we are searching for a Calabi-Yau metric on the singular 
conifold with an $SU(2)\times U(1)$ symmetry, and the discrete, $\ZZ_2$ $\cR$ 
symmetries described in \KlebanovHH.    The metric must agree with the 
more standard metric \CandelasJS\ for small values of the radial coordinate, 
and it must reduce to a flat metric on $\IC^2/\ZZ_2 \times \IC$ in 
limit of some large appropriately chosen radial coordinate.

\newsec{The conifold,  its complex structure and hermitian metrics}

\subsec{Describing the conifold}
 
As explained in \CandelasJS, it is very convenient to describe the 
conifold  in terms of the matrix:
\eqn\Wmatrix{\cW ~\equiv~ {1 \over \sqrt{2}} \, \left(\matrix{w_3 + i w_4& 
w_1 - i w_2 \cr  w_1 +  i w_2 &- w_3 + i w_4}\right) \,,}
The vanishing of the determinant of this matrix defines the conifold:
\eqn\deteqn{ {\rm det} (\cW) ~=~ -\coeff{1}{2}\,( w_1^2 ~+~ w_2^2 ~+~ 
w_3^2 ~+~   w_4^2) ~=~ 0 \,,}
while the obvious norm defines a natural radial coordinate:
 \eqn\normW{ {\rm Tr} ( \cW^\dagger\, \cW ) ~=~   |w_1|^2 ~+~ |w_2|^2 ~+~ 
 |w_3|^2 ~+~   |w_4|^2  ~\equiv~ r^2 \,.}
 The surface of constant $r$ is the  $T^{(1,1)}$ space:
 \eqn\Toneone{T^{(1,1)}~=~ {SU(2) \times SU(2) \over U(1)} \,,}
which can be  given a Einstein metric  \RomansAN.  Taking the cone
 over this leads to the well-known Ricci-flat metric on the conifold 
 \refs{ \CandelasJS, \KlebanovHH}.  This metric has  
 $SU(2)_L \times SU(2)_R \times U(1)$ invariance where the $SU(2)$'s act 
 on the left and right of $\cW$, while the $U(1)$ is an overall phase rotation
 on the $w_j$.
 
As described  in the previous section, we seek a broader class of metrics
on the conifold:  Metrics that preserve the complex structure
and have an  $SU(2) \times U(1)$  symmetry where the 
$U(1)$ is the phase rotation on the $w_j$ and the $SU(2)$ is
the diagonal subgroup of $SU(2)_L \times SU(2)_R$.   Note that this 
particular $SU(2)$ leaves $w_4$ invariant, while transforming $(w_1,w_2,w_3)$
as a triplet.   Thus the continuous symmetry acts as $SO(3) \times U(1)$.

Also note that if one uses the identification
of coordinates and fields in \gaugeinvvar\ and \wfromz\ then the
discrete symmetry, \intersymm, of the field theory becomes the 
reflection: 
\eqn\refsymm{  w_1 \to -w_1 \,, \quad w_2 \to -w_2 \,, \quad w_3 \to -w_3 \,, \quad 
w_4 \to w_4 \,.}
Thus we are actually looking for an $O(3) \times U(1)$ invariant
metric.

\subsec{Breaking some symmetry and a degenerate limit}

Consider the ``trivial'' deformation of the conifold:
\eqn\defconeqn {w_1^2~+~w_2^2 ~+~ w_3^2~+~ \varepsilon^2 \, w_4^2  ~=~ 0\,. }
As we discussed earlier, the parameter, $\varepsilon$,  can be removed by
rescaling $w_4$.  However, once one has chosen a metric on the conifold, this
scale parameter has meaning, and breaks the symmetry to  
$O(3) \times U(1)$.  Moreover, if one takes the
degenerate limit $\varepsilon=0$ this modifies the complex structure
and enhances the symmetry to $SU(2) \times U(1) \times U(1)$.  
In particular, the underlying complex manifold becomes the orbifold 
$\IC^2/\ZZ_2 \times \IC$.  The single factor of $\IC$ is simply parametrized
by $w_4$, while the $\IC^2$ can be parametrized by two coordinates
$(\zeta_1, \zeta_2)$ that are identified under $\zeta_j \to -\zeta_j$.  The 
invarants under this action are:
\eqn\Ztwoact{ w_1 ~=~ \coeff{1}{2}\, (\zeta_1^2+ \zeta_2^2) \,, \qquad 
w_2 ~=~  \coeff{i}{2}\, (\zeta_1^2 - \zeta_2^2)\,, \qquad 
w_3 ~=~ i\, \zeta_1\, \zeta_2 \,,  }
and these satisfy \defconeqn\ with $\varepsilon =0$.  Note that
under the residual $SU(2)$ symmetry $(\zeta_1, \zeta_2)$ 
transforms as a doublet.  The two $U(1)$ symmetries are
the now-independent phase rotations of $(\zeta_1, \zeta_2)$
and $w_4$.    

Note that in this degeneration limit we have recovered the 
underlying manifold and symmetry structure of the $\ZZ_2$ orbifold  of 
the solutions considered in   \refs{\PilchEJ,\PilchFU}.
We will discuss this further in section 6.

\subsec{Some useful coordinates}

We wish to parametrize the conifold in terms of the residual
$SU(2) \times U(1)$ symmetry.  To this end, introduce:
\eqn\Wzero{\cW_0 ~\equiv~ {1 \over \sqrt{2}}\,\sqrt{\mu^2 - \nu^2} \, \oneone ~+~ 
{1 \over \sqrt{2}}\,\left(\matrix{0 &  \mu+\nu  \cr  \mu-\nu  & 0}\right) \,,}
where $\mu, \nu$ and $\phi$ are real, with $|\mu| \ge  |\nu|$.
A general matrix of the form \Wmatrix\ can then be obtained by 
conjugating with a single matrix, $\cL \in SU(2)$, and multiplying 
by a phase, $e^{i\phi} \in U(1)$ :
\eqn\conjact{\cW  ~=~ e^{i\phi}\,\cL\, \cW_0 \, \cL^\dagger \,.}
One can see that this is possible by considering the real
and imaginary parts of $w_j = x_j + i y_j$, and denoting the
first three components by $3$-vectors $\vec x$ and $\vec y$.    
First use the overall phase to set $x_4 =0$, and then the imaginary
part of the conifold equation \deteqn\ means that $\vec x \cdot \vec y =0$.
One can now use all of the $SO(3)$ symmetry to put these orthogonal
vectors in the form $\vec x = (\mu,0,0)$ and $\vec y = (0,\nu,0)$ for some
$\mu, \nu > 0$.  The real part of the conifold equation
implies $\mu^2 = \nu^2 +y_4^2$, and hence $\mu \ge \nu$. We thus obtain
$w_1 =\mu$, $w_2 = i \nu$, $w_3 = 0$ and $w_4 =  i \sqrt{\mu^2 - \nu^2}$,
which is \Wzero, provided that we let
$\mu$ and $\nu$ be both positive or negative with $|\mu| \ge  |\nu|$.

We parametrize the $SU(2)$ matrices in terms of Euler angles.  That is, we write
\eqn\Lform{\cL ~=~ \cR_{12}(\varphi_2) \, \cR_{13}(\varphi_1) \,\cR_{12}(\varphi_3) \,,}
where
\eqn\Euler{ \cR_{12}(\varphi )  ~\equiv ~  \left(\matrix{e^{-{i \over 2}\, \varphi} & 
0 \cr  0  & e^{+ {i \over 2}\, \varphi} } \right)    \,, \qquad  \cR_{13}(\varphi )  ~\equiv ~  
\left(\matrix{\cos({1\over 2}\, \varphi) & - \sin({1\over 2}\, \varphi) \cr  
\sin({1\over 2}\, \varphi)  & \cos({1\over 2}\, \varphi)}\right) \,.}
It is convenient to introduce the invariant $1$-forms, $\sigma_j$ where:
\eqn\sigdef{ \cL^\dagger\, d \cL ~=~   {1 \over 2}\, \left( \matrix{ i \, \sigma_3 & 
 \sigma_1 - i \, \sigma_2 \cr  -( \sigma_1 + i \, \sigma_2)  &  -i \, \sigma_3 } \right)   \,.}
Explicitly, one has:
\eqn\oneforms{\eqalign{\sigma_1 ~\equiv~&  \cos \varphi_3\, d\varphi_1 ~+~ 
\sin\varphi_3\, \sin\varphi_1\, d \varphi_2 \,, \cr
\sigma_2 ~\equiv ~&  \sin\varphi_3\, d\varphi_1 ~-~ 
\cos\varphi_3\, \sin\varphi_1\, d \varphi_2 \,, \cr
\sigma_3 ~\equiv ~&  \cos\varphi_1\, d\varphi_2 ~+~   d \varphi_3  \,.}}

For future reference we note that in these coordinates, the degenerate
limit that takes us from the conifold to the orbifold $\IC^2/\ZZ_2 \times \IC$
is to take $|\mu|, |\nu| \to \infty$ with $\mu - \nu $ finite, or $|\mu|, |\nu| \to \infty$ 
with  $\mu + \nu $ finite.

\subsec{The $(1,0)$-forms}

Since the conifold is a $3$-fold, there are three independent $(1,0)$-forms
at each point, and it turns out there there is a $SU(2) \times U(1)$
invariant basis for these differentials.  In terms of complex coordinates,
these are $\sum_{j=1}^3 \bar z_j d z_j$,  $\bar z_4 d z_4$ and 
$\bar z_4 ( \epsilon^{ijk} \bar z_i z_j d z_k)$, while in terms of the 
coordinates introduced above, these may be reduced to the equivalent basis:
\eqn\holforms{\eqalign{\omega_1 ~\equiv~ & d \mu ~+~ i (\mu \, d \phi ~+~ \nu \, 
\sigma_3) \,, \cr 
\omega_2  ~\equiv~ & d \nu ~+~ i (\nu \, d \phi ~+~ \mu \, \sigma_3) \,, \cr
\omega_3  ~\equiv~ &  \mu \, \sigma_1 ~+~ i \, \nu \, \sigma_2  \,.}}

It is instructive to note that the {\it degenerate}  conifold also has the 
$SU(2)$ invariant $(1,0)$-form:
\eqn\degenholform{ \omega_0  ~\equiv~   \zeta_2 \, d \zeta_1 
-  \zeta_1 \, d \zeta_2  ~\sim~ (\sigma_1 ~+~ i  \, \sigma_2) ~=~
{1 \over 2\, u\, v}\, \big( (\mu+\nu)\, \omega_3 + (\mu-\nu)\, \overline{ \omega}_3 \big) \,. }
Observe that while this is a $(1,0)$-form in the 
complex structure of the degenerate conifold, it is not of type $(1,0)$
in the complex structure of the non-degenerate conifold.  Thus the limit
$\varepsilon \to 0$ in \defconeqn\ involves a discontinuity in the
complex structure.

\subsec{Hermitian metrics}

The most general hermitian metric on the conifold can be written in the form
\eqn\hermmet{
ds^2  ~=~ g_{i\, \bar \jmath} ~\omega^i \, \bar \omega^{\bar \jmath} \,,}
for some metric coefficients, $g_{i\, \bar \jmath}$, that are arbitrary
functions of the coordinates.
The $SU(2) \times SU(2) \times U(1)$ invariant, Ricci-flat metric
on the conifold is:
\eqn\rndmet{\eqalign{ds^2 ~=~& \coeff{1}{3}\, u^{-{2 \over 3}} \, \Big(  
 \coeff{1}{3}\,  |\omega_1|^2 ~+~    |\omega_2|^2  ~+~    |\omega_3|^2 ~+~ 
 {1\over (\mu^2 - \nu^2)}\,  |\mu\, \omega_1-\nu\, \omega_2|^2  \Big) \cr ~=~& 
d\rho^2 ~+~   \coeff{1}{3}\,  \rho^2  \, \Big( d\theta^2 ~+~ \sigma_1^2~+~\cos^2 \theta\,
 \sigma_2^2 ~+~\sin^2 \theta\, \sigma_3^2  ~+~\coeff{4}{3}\, (d\phi  +  \cos \theta\, 
  \sigma_3)^2\, \Big) \,. }}
where $\mu = \rho^{3/2}$ and $\nu = \rho^{3/2} \cos\theta$.
This is the cone over $T^{(1,1)}$, but parametrized in a slightly unusual
manner.  Remember that we want to preserve and make manifest the 
diagonal $SU(2)$ in $SU(2)_L \times  SU(2)_R$.  If one does this 
on the $S^2 \times S^2$ base of the $T^{(1,1)}$ space then one
can write the standard metric on this product space as:
\eqn\prodmet{\eqalign{ds_4^2 ~=~& \big(d\theta_1^2 ~+~ \sin^2\theta_1\, 
d\phi_1^2\big) ~+~ \big(d\theta_2^2 ~+~ \sin^2\theta_2\, 
d\phi_2^2\big)  \cr ~=~&  2\, \big( d\theta^2 ~+~ \sigma_1^2~+~\cos^2 \theta\,
 \sigma_2^2 ~+~\sin^2 \theta\, \sigma_3^2   \big) \,. }}
Since $d( 2\, \cos \theta\,  \sigma_3)$ yields a K\"ahler form on 
\prodmet, we see that \rndmet\ involves the proper Hopf fibration 
to make $T^{(1,1)}$.

Imposing the condition that the metric be $SO(3) \times U(1)$ invariant implies
that  the  metric can only depend explicitly upon $\mu$ and $\nu$.  It is also
easy to see that the $\ZZ_2$ symmetry:  
\eqn\extrasymm{ \varphi_3 \to \varphi_3 ~+~ \pi \qquad
\Rightarrow \qquad \sigma_1 \to - \sigma_1\,, \ \ \ \sigma_2 \to - \sigma_2 }
is exactly that of \refsymm.  This is because $w_4 = -i \sqrt{\mu^2 - \nu^2} e^{i \phi}$
is invariant under the shift of $\varphi_3$, while in the conjugation by 
${\cal L}$, this shift has the effect of negating all the entries in the second
matrix of \Wzero. (In section 5  we will give explicit formulae
for the $w_j$, from which this result will be obvious.)

Requiring that the metric be invariant under this $\ZZ_2$ symmetry  has the effect of 
requiring that $\omega_3$ only appear paired  with $\bar \omega_3$ in \hermmet.  
 The metric thus has the form:
\eqn\symmform{ds^2  ~=~ Q_1 \, d\mu^2  + Q_2 \,  d\nu^2 + Q_3 \,  d\mu\, d\nu + 
Q_4 \, |\omega_3|^2 + Q_5 \, (\sigma_3 + Q_6 \, d\phi)^2 ~+~ Q_7 \, d\phi^2 \,,}
for some arbitrary functions, $Q_j(\mu,\nu)$.  By introducing a new
coordinate, $u(\mu,\nu)$, we can convert the $(\mu,\phi)$ part of this
metric to $\hat Q_1 \, (du^2 + u^2 d \phi^2)$, for some new function
$\hat Q_1$.  By introducing a new coordinate, $v(u,\nu)$, we
can then eliminate the $du\, d\nu$ term.  This then leads to our final
metric Ansatz:
\eqn\metAnsatz{ds^2  ~=~ H_1^2\,( du^2  + u^2 \, d\phi^2) ~+~ 
H_2^2\, dv^2  ~+~  H_3^2\, \sigma_1^2  ~+~  H_4^2\, \sigma_2^2  ~+~  
H_5^2\, v^2\, (\sigma_3 + H_6\, d\phi)^2 \,,}
for some arbitrary functions, $H_j(u,v)$.   The existence of the complex structure 
with respect to which the metric \metAnsatz\ is hermitian
imposes relations upon the $H_j$, but we will not make this assumption, and instead 
recover it from the supersymmetry variations.  Our starting point will therefore be this
metric Ansatz. However we should mention, that a $u$-independent re-definition of $v$ does
not change the form of this Ansatz. We will later see how to choose a good gauge for this.

Note that this metric Ansatz is a simple generalization of the one employed in 
\PilchYG.

\newsec{The Calabi-Yau metric from the Killing spinors}

The  traditional way  to obtain a ``simple'' equation the Ricci-flat metric on the 
a K\"ahler manifold is to write down the Ricci-flat  condition in terms of the 
K\"ahler potential.  This generates a  Monge-Amp\`ere 
equation, which, for a $3$-fold is a sixth-order, non-linear PDE.  There
are existence theorems for solutions of this equation \refs{\tianyaua,\tianyaub},
and these will be discussed later.

The problem we wish to solve here has some extra global
symmetry, and we will use the supersymmetry  in an approach that closely follows that of 
\refs{\GowdigereJF,\PilchJG,\PilchYG,\NemeschanskyYH}.
That is, we start with the metric Ansatz \metAnsatz, make 
an algebraic prescription for the supersymmetry bundle, and
go on to derive all the essential equations from the supersymmetry
variations.   The result is significantly simpler than the direct
output of the Monge-Amp\`ere equation.

\subsec{Solving the Ansatz}

We begin by introducing the frames:
\eqn\frames{\eqalign{e^1 ~\equiv~ & H_1\, du\,, \qquad e^2 ~\equiv~H_2\, dv\,, \qquad 
e^3 ~\equiv~H_3\, \sigma_1 \,, \qquad e^4~\equiv~ H_4\, \sigma_2 \,, \cr 
e^5 ~\equiv~  & H_5\,v\, (\sigma_3 + H_6\, d\phi)  \,, \qquad 
e^6 ~\equiv~H_1\, u\,d\phi\,.}}
This metric has a natural almost-complex structure (indeed, it is the one 
inherited from the conifold, and will ultimately become  a K\"ahler form):
\eqn\acstr{J ~=~  e^1 \wedge e^6 ~+~   e^3 \wedge e^4 ~+~e^2 \wedge e^5   \,.}
Based upon this, we introduce the helicity projectors:
\eqn\projs{\Pi_{1}~=~  \coeff{1}{2}\big(\oneone -i \,\Gamma^{1}\Gamma^{6} \big)\,,
\qquad \Pi_{2}~=~  \coeff{1}{2}\big(\oneone - i\, \Gamma^{3}\Gamma^{4} \big)\,,
 \qquad \Pi_{3}~=~  \coeff{1}{2}\big(\oneone - i \,\Gamma^{2}\Gamma^{5} \big)\,,}
and look for covariant constant spinors that satisfy:
\eqn\projconds{\Pi_{j}\, \epsilon ~=~ 0 \,, \qquad j = 1,2,3\,.}
There will then be a second covariant constant spinor that can be 
obtained by complex conjugation and whose helicities are thus
exactly the opposite of those in \projs,\projconds.

In order to solve 
\eqn\covconst{\nabla_\mu \epsilon ~=~ 0 \,,}
it is helps to fix the dependence on the coordinates 
as far as possible before analyzing the equations in detail.
The dependence upon angular coordinates can be obtained 
using the symmetries and the Lie derivative on spinors
\eqn\LieDeriv{L_{K}\, \epsilon ~\equiv~ K^\mu \, \nabla_\mu \, \epsilon
~+~ \coeff{1}{4}\, \nabla_{[\mu} K_{\nu]} \, \Gamma^{\mu \nu} \, \epsilon \,.}
The covariant constant spinors must be singlets under the $SU(2)$ action,
and so the corresponding $L_{K}\, \epsilon $ must vanish.  With the frames
used above, this implies that $\epsilon$ is independent of the $\varphi_j$.
The covariant constant spinors can be charged under the $U(1)$, and this charge
will be correlated with helicity.  When one simplifies the corresponding Lie derivative,
one finds that this amounts to having:
\eqn\epschg{\partial_\phi \epsilon ~=~i\,q\, \epsilon \,,}
for some constant charge, $q$.  However, by taking the flat space limit
of \frames\ one can fix 
$$
q= {1 \over 2} \,,
$$
provided that $\phi$ is normalized to the periodic range: $0 \le \phi <  2\pi$.
Finally, one can fix the normalization of $\epsilon$ by using the fact that
\covconst\ implies that $\bar \epsilon \,\Gamma^\mu \epsilon$ is constant.  This 
means that $\epsilon$ can be chosen to be independent of $u$ and $v$.

Substituting all of this information into  \covconst\
yields a fairly involved system of equations that is, in fact, relatively
easy to disentangle.  One finds the following simple conditions:
\eqn\firstone{  H_5 ~=~  c_0\,  H_2^{-1} \,, }
\eqn\secondone{H_4 ~=~
(\coeff{1}{2}\, c_0\,   v^2 + c_1)\, H_3^{-1} \,,  }
\eqn\firstset{ u\,\partial_u H_2^2 ~+~ c_0 \,  v\,\partial_v H_6 ~=~  0 \,, \qquad
u\,\partial_v H_1^2 ~-~ c_0  \, v\,\partial_u H_6 ~=~   0 \,, }
 where   $c_0$ and $c_1$ are constants of integration.  One also
obtains some more complicated conditions that interrelate the
derivatives of $H_3$, $H_1$ and $H_2$.  The latter can be simplified
and integrated by making the substitution:
\eqn\secondset{ H_3^2  ~=~   (\coeff{1}{2}\, c_0\,   v^2 +  c_1) \,
 \left( 1 + e^{2 \, h} \over   1 - e^{2 \, h} \right) \,, \qquad  
 H_4^2 ~=~   (\coeff{1}{2}\, c_0\,   v^2 +  c_1) \,
  \left( 1 -  e^{2 \, h} \over   1 + e^{2 \, h} \right) \,, }
for some function $h(u,v)$.  In making this substitution we have used
\secondone.  One then finds that the equations 
for $H_1$ and $H_2$ can be integrated to give:
\eqn\thirdset{ H_1^2  ~=~   {c_0   \, m(u)  \over  v \,
 ( c_0\,   v^2 + 2\, c_1)} \, \sinh(2\,h) \, \partial_v h \,, \qquad
  H_2^2  ~=~   - c_0\, v\, \partial_v h \,, }
for some, as yet,  arbitrary function 
$m(u)$.   Using this in the remaining supersymmetry variations one
finds that:
\eqn\lastone{H_6~=~ u\, \partial_u h \,,}
and thus the last equation in \firstset\ yields:
\eqn\firstmastereqn{ {1 \over  u} \, \partial_u \big(u\, \partial_u h \big) ~-~
{1  \over v} \,\partial_v \bigg({m(u) \over v \,  ( c_0\,   v^2 + 2\, c_1)}\,
\sinh(2\,h) \, \partial_v h \bigg) ~=~ 0\,.}
Finally, the supersymmetry in the $\phi$-direction yields the condition
\eqn\setcharge{\partial_\phi \epsilon ~=~ i
\,\Big[  \,{1 \over 2} + {m'(u) \over 4 \, m(u) }\, \Big] \epsilon \,,}
we  thus find that 
\eqn\mfunction{ m(u) ~=~ c_2\, u^\alpha  \qquad  \Rightarrow \qquad
q ~=~\Big[  \,{1 \over 2} + {\alpha \over 4  }\, \Big]  \,,}
for some constants $c_2$ and $\alpha$.  We thus obtain
\eqn\mastereqn{ {1 \over  u^{\alpha+1} } \, \partial_u \big(u\, \partial_u h \big) ~-~
{c_2 \over v} \,\partial_v \bigg({1\over v \,  ( c_0\,   v^2 + 2\, c_1)}\,
\sinh(2\,h) \, \partial_v h \bigg) ~=~ 0\,.}

This ``master equation'' determines the complete solution to our problem:
Given a solution to this equation, one can determine the metric functions 
$H_j$ from \thirdset,  \secondset, \firstone\ and \lastone.  One then finds
that one does indeed have a covariantly constant spinor of the form described
above, and one can explicitly verify that the metric is Ricci-flat.   Indeed,
the form of the function $m(u)$ in \mfunction\ follows directly from
setting the Ricci tensor to zero.  

The ``master equation,'' while non-linear, is actually a relatively simple,
quasi-linear, second order PDE.  It is certainly simpler than the Monge-Amp\`ere
equation, and more significantly for our work, it is very similar to
the classes of PDE's found in \refs{\GowdigereJF,\PilchJG,\PilchYG,\NemeschanskyYH}.

\subsec{Conical singularities and the period of $\phi$}

If we have a Ricci-flat metric of the form \metAnsatz\ then it is still, at least
locally, Ricci-flat if one changes the period of $\phi$ to any value one desires.
The potential cost of this is the introduction of
a conical singularity, but the details will depend upon
the asymptotics of $H_1$.  If $H_1 \to 1$ then the non-conical 
choice is $0 \le \phi < 2 \pi$.  

The fact that one can choose the period of $\phi$ has implicitly appeared in our 
analysis above via the constant parameter $\alpha$.  Specifically, a change
of variable $u =w^\gamma$ combined with $\phi = \gamma \psi$ is
an analytic change of variable:  $z \equiv u e^{i \phi} \to z^\gamma$, and it 
conformally maps the $(u,\phi)$-metric via:
\eqn\uphimet{du^2 +  u^2 \,d \phi^2 \to w^{2(\gamma-1)} \, \big(
dw^2 +  w^2 \,d \psi^2 \big) \,.}
The net result is that \metAnsatz\ keeps the same form, except
that the period of $\phi$ is scaled by $\gamma$.  In the foregoing
analysis this change of variable merely generates a constant 
scale in $u \partial_u$, but sends ${1 \over u } \partial_u$
to ${1 \over \gamma  w^{(2 \gamma-1)}} \partial_w$ which means
that such a re-definition can be used to set $\alpha = 0$ in
\mastereqn.  While this might be a convenient choice, one must remember
that this choice imposes a particular periodicity upon the $\phi$-coordinate,
and in practice one may want to preserve the free parameter $\alpha$.

\newsec{The complex geometry}

In order to understand the geometric significance of the solution 
it is instructive  to re-derive them by direct analysis of the holonomy.

\subsec{The Calabi-Yau conditions}

The covariantly constant spinor defines an almost complex 
structure that can be expressed by choosing the hermitian vielbein
\eqn\hermframes{\eqalign{
E^1 ~\equiv~ & H_1\,(du+iu\,d\phi)\,, \cr
E^2 ~\equiv~ & H_2\,dv+i H_5\,v\, (\sigma_3 + H_5\, d\phi)\,, \cr
E^3 ~\equiv~ & H_3\,\sigma_1+iH_4\,\sigma_2\,.}}
This vielbein defines the K\"ahler form $J=\sum_iE^i\wedge E^{\bar \imath}$ and a 
hermitian metric.

The spin connection can now be expressed in terms of holomorphic and 
anti-holomorphic indices. In order for the spin connection to have $U(3)$
holonomy, it has to be a one-form with values in the Lie algebra, ${\cal U}(3)$.
This is achieved by requiring the spin connection to be of the form
$\omega^i{}_j$ or, equivalently, $\omega_{\bar \imath j}$.

Since $\omega^i{}_{\bar \jmath}\equiv 0$, the torsion 2-form has the following form
\eqn\torsionform{T^i ~\equiv~ dE^i+\omega^i{}_j\,\wedge\,E^j\,.}
This can be decomposed into a $(2,0)$-form, a $(1,1)$-form and a $(0,2)$-form
part. We will step by step require the vanishing of all these torsion parts.

First, consider the $(0,2)$-form part of the torsion:
\eqn\torsiona{(T^i)_{(0,2)}=(dE^i)_{(0,2)}\,.}
Requiring that this part of the torsion vanish is the same as requiring that the 
complex structure be integrable. For the Ansatz \hermframes\ the only 
non-trivial components of the $(0,2)$-form part of the torsion are 
$T^2{}_{\bar 1\bar 2}$, $T^3{}_{\bar 1\bar 3}$ and $T^3{}_{\bar 2\bar 3}$.
This leads to the equations
\eqn\cxstrintegr{
\partial_u\left({H_2 \over vH_5}\right)+\partial_v\left({H_6 \over u}\right)=0\,,\quad
\partial_u\log{{H_3 \over H_4}-1 \over {H_3 \over H_4}+1}={2H_6 \over u}\quad{\rm and}
\quad
\partial_v\log{{H_3 \over H_4}-1 \over {H_3 \over H_4}+1}=-{2H_2 \over vH_5}\,.}
The first equation is an integrability condition and says, that there is a function $h$,
such that
\eqn\defineh{H_6=u\partial_uh \quad{\rm and}\quad {H_2 \over H_5}=-v\partial_vh\,.}
The other two equations then imply, that 
\eqn\athreeoverafour{{H_3 \over H_4}=-\coth(h+c)\,,}
where $c$ is an integration constant, which can be absorbed into the definition
of $h$. However, there is a second solution with $H_3=\pm H_4$, which corresponds 
to $c \to \pm\infty$. For now we will focus on the nondegenerate case. The Hermitean
frame then reduces to
\eqn\cxframes{\eqalign{
E^1 ~=~ & H_1\,(du+iu\,d\phi)\,, \cr
E^2 ~=~ & H_2\,\left(dv-{i \over \partial_vh}\,(\sigma_3 + u\,\partial_uh\, d\phi)\right)\,, \cr
E^3 ~=~ & H_3\,(\sigma_1-i\coth(h)\,\sigma_2)\,,}}
with the elementary $(1,0)$-forms
\eqn\newholforms{\eqalign{
\omega^1 ~\equiv~ & du+iu\,d\phi\,, \cr
\omega^2 ~\equiv~ & dv-{i \over \partial_vh}\,(\sigma_3 + u\,\partial_uh\, d\phi)\,, \cr
\omega^3 ~\equiv~ & \sinh(h)\,\sigma_1-i\cosh(h)\,\sigma_2\,.}}

Next, the $(1,1)$-form part of the torsion is:
\eqn\torsionb{(T^i)_{(1,1)}=(dE^i)_{(1,1)}+\omega_{\bar k}{}^i{}_j\,E^{\bar k}\,\wedge\,E^j\,.}
The vanishing of this part of the torison allows one to read off the non-trivial part of the
spin connection immediately:
\eqn\holospinconn{\eqalign{
&\omega_{\bar 1\bar 1 1}=-{1 \over 2H_1}\partial_u\log(uH_1)\,,\quad
\omega_{\bar 2\bar 1 1}=-{1 \over 2H_1}\partial_v\log H_1\,,\quad
\omega_{\bar 1\bar 2 1}=-{vH_5\partial_uH_6 \over 2uH_1^2}\,,\cr
&\omega_{\bar 2\bar 2 1}={1 \over 4H_1}\left(\partial_u\log{H_2 \over H_5}-{vH_5
\partial_vH_6 \over uH_2}\right)\,,\quad
\omega_{\bar 1\bar 2 2}=-{1 \over 4H_1}\left(\partial_u\log(H_2H_5)+
{vH_5\partial_vH_6 \over uH_2}\right)\,,\cr
&\omega_{\bar 2\bar 2 2}=-{1 \over 2H_2}\partial_v\log(vH_5)\,,\quad
\omega_{\bar 3\bar 2 3}=-{vH_5 \over 2H_3H_4}\,,\cr
&\omega_{\bar 3\bar 3 1}={1 \over 4H_1}\left(\partial_u\log{H_3 \over H_4}+
{H_6 \over u}\left({H_3 \over H_4}-{H_4 \over H_3}\right)\right)\,,\cr
&\omega_{\bar 3\bar 3 2}={1 \over 4H_2}\left(\partial_v\log{H_3 \over H_4}-
{H_2 \over vH_5}\left({H_3 \over H_4}-{H_4 \over H_3}\right)\right)\,,\cr
&\omega_{\bar 1\bar 3 3}=-{1 \over 4H_1}\left(\partial_u\log(H_3H_4)+
{H_6 \over u}\left({H_3 \over H_4}+{H_4 \over H_3}\right)\right)\,,\cr
&\omega_{\bar 2\bar 3 3}={1 \over 4H_2}\left(-\partial_v\log(H_3H_4)+
{H_2 \over vH_5}\left({H_3 \over H_4}+{H_4 \over H_3}\right)\right)\,.}}
One can use the reality relation, $\omega_{k \bar \imath j}=-(\omega_{\bar k\bar \jmath i})^*$,
to determine the remainder of the spin connection.

The the spin connection can now be inserted into the equation for the vanishing of the
$(2,0)$-form part of the torsion. This is equivalent to the K\"ahler condition $dJ=0$, 
and reduces to four non-trivial equations:
\eqn\kahlercona{\eqalign{
&\partial_v(H_1^2)={v \over u} H_2H_5\partial_uH_6\,,\cr
&\partial_u(H_2H_5)=0\,,\quad
\partial_v(H_3H_4)=vH_2H_5\quad{\rm and}\quad
\partial_u(H_3H_4)=0\,.
}}

The last three equations show  that $H_3H_4$ can be an arbitrary function, $f(v)$, 
and that $H_2H_5$ is given by $H_2H_5 = {1 \over v} \partial_v f(v)$.  Fixing $f(v)$ 
fixes the freedom to re-define $v$ by an arbitrary function of $\tilde v(v)$. 
Using \defineh\ and \athreeoverafour\ , we get the relations
\eqn\kahlerconb{H_3^2=f(v)\coth(h)\quad{\rm and}\quad H_2^2=\partial_vh\,\partial_vf\,.}
The first condition \kahlercona\ then determines $H_1^2$ up to a function of $u$ 
only.

Finally, we have the condition for $SU(3)$ holonomy, as opposed to $U(3)$ holonomy. 
From the spin connection
we can extract a $U(1)$ connection $\tilde\omega=\omega^i{}_i$. Requiring that 
$\tilde\omega$ is a flat connection is now equivalent to $\omega$ having $SU(3)$ 
holonomy. This condition is $d\tilde\omega=0$ or $\tilde\omega=d\lambda$.
Using \holospinconn, we see that
\eqn\uoneconn{\tilde\omega=
{iu \over 2}\,\partial_u\log{u^2H_1^2H_2^2H_3^2 \over (\partial_vh)^2 \cosh^2h}\,d\phi-
{i \over 2\partial_vh}\,\partial_v\log{u^2H_1^2H_2^2H_3^2 \over (\partial_vh)^2 \cosh^2h}\,(\sigma_3+u\partial_uh\,d\phi)\,.}
A way to make this a total derivative is by requiring, that 
\eqn\sucond{{u^2H_1^2H_2^2H_3^2 \over (\partial_vh)^2 \cosh^2h}=c_2u^{\alpha+2}\,.}
Inserting this together with \kahlerconb\ into the first equation of \kahlercona\ gives
the master equation
\eqn\mastereqnb{{c_2 \over \partial_vf}\,\partial_v\left({1 \over \partial_v(f^2)}\,\sinh(2h)\,\partial_vh\right)+{1 \over u^{\alpha+1}}\,\partial_u(u\partial_uh)=0\,,}
where $c_2$ can be absorbed into $f$. Choosing $f$ properly, one can easily 
recover \mastereqn.

The Calabi-Yau condition can also be formulated in terms of a holomorphic 
(3,0)-form
\eqn\Omdefn{\Omega ~\equiv~e^{i \phi} E^1 \wedge  E^2 \wedge E^3 \,,}
where the phase dependence can be fixed by requiring the proper transformation
properties under the $\cR$-symmetry.
To have a Calabi-Yau manifold this form must be holomorphic, and satisfy 
\eqn\formcond{\Omega \wedge \overline \Omega ~=~{1 \over 3!}\, J \wedge  J \wedge J  \,.}
The latter is trivially satisfied, while the former is equivalent to requiring
that $\Omega$ be closed. This leads again to \mastereqnb.

In general it is easy in the non-coordinate base to impose \formcond. But then 
it is harder to impose that $J$ and $\Omega$ are closed. On the other hand
in a coordinate base it is easy to impose that $J$ and $\Omega$ are closed,
but then it is hard to impose \formcond. The true art is to impose both at the 
same time.

\subsec{Holomorphic coordinates}

Finally, it is instructive to see how the conifold is parametrized in terms
of the new variables, $u$ and $v$.  To determine the relationship
one notes that \holforms\ and \newholforms\ must be
linear combinations of one another.  It is trivial to solve this 
system and thereby obtain:
\eqn\holformrelns{\mu ~=~  u\, \cosh(h) \,, \qquad \nu ~=~  u\, \sinh(h)   \,.}
One can then compute the holomorphic coordinates explicitly using
\Wzero\ and \conjact\ to obtain: 
\eqn\wonetwo{\eqalign{w_1  ~=~  & \coeff{1}{2}\,(i w_4) \, \big[ \cos^2(\coeff{1}{2} \varphi_1)
\, (\xi\, e^{i \varphi_2} +(\xi \, e^{i \varphi_2})^{-1} )   ~-~ \sin^2(\coeff{1}{2} 
\varphi_1 ) \, (\xi\, e^{- i \varphi_2}  +(\xi \, e^{- i \varphi_2})^{-1}) \big] \cr 
w_2  ~=~  & \coeff{i}{2}\, (i w_4) \, \big[ \cos^2(\coeff{1}{2} \varphi_1)  \,
(\xi\, e^{i \varphi_2} - (\xi \, e^{i \varphi_2})^{-1} )   ~+~  \sin^2(\coeff{1}{2} 
\varphi_1 ) \, (\xi\, e^{-i \varphi_2}  -(\xi \, e^{- i \varphi_2})^{-1})  \big]  \,,}}
and
\eqn\wthreefour{w_3 ~=~ \coeff{1}{2}\,(i w_4) \, (\xi+\xi^{-1}) \, \sin \varphi_1  \,, 
\qquad w_4 ~=~ - i u\,e^{i \phi}\,,}
where
\eqn\xidefn{\xi~\equiv~ e^{h + i\varphi_3} \,.}

One can naturally parametrize $\IC^2$ using the $SU(2)$ matrix $\cL$
 and a radial coordinate, $r$, and one can obtain:
 \eqn\zetaonetwo{\zeta_1 ~=~ r \,  \cos(\coeff{1}{2} \varphi_1)  \,e^{{i \over 2}\,
 (\varphi_2 + \varphi_3)} \,, \qquad \zeta_2 ~=~ i\, r \, \sin(\coeff{1}{2} \varphi_1)  \,
 e^{-{i \over 2}\,  (\varphi_2 - \varphi_3)} \,.}
 Taking $r = e^{h/2}$, we see that \wonetwo\ and \wthreefour\ may be rewritten as:
\eqn\wonetwo{\eqalign{w_1  ~=~  & \coeff{1}{2}\,(i w_4) \, \big[ (\zeta_1^2 + \zeta_2^2)
~+~ e^{-2\,h}\, (\bar \zeta_1^2 + \bar \zeta_2^2)  \big] \,, \cr 
w_2  ~=~  & \coeff{i}{2}\, (i w_4) \, \big[ (\zeta_1^2 - \zeta_2^2)
~-~ e^{-2\,h}\, (\bar \zeta_1^2 - \bar \zeta_2^2)  \big] \,,\cr 
w_3 ~=~ &- \coeff{i}{2}\,(i w_4) \, (\zeta_1\, \zeta_2  ~+~ e^{-2\,h}\,  \bar \zeta_1\, 
\bar \zeta_2 )  \,,  \qquad w_4 ~=~ - i u\,e^{i \phi}\,.}}
This shows how we are writing the conifold as a deformation of
the orbifold \Ztwoact, and that the confold and orbifold have distinct
complex structures.  

Alternatively, one can take:
 \eqn\zetaonetwo{\zeta_1 ~=~ r \,  \cos(\coeff{1}{2} \varphi_1)  \,e^{-{i \over 2}\,
 (\varphi_3 + \varphi_2)} \,, \qquad \zeta_2 ~=~ i\, r \, \sin(\coeff{1}{2} \varphi_1)  \,
 e^{-{i \over 2}\,  (\varphi_3 - \varphi_2)} \,.}
and set   $r = e^{- h/2}$.  One then finds that \wonetwo\ and \wthreefour\ may be rewritten as:
\eqn\wonetwo{\eqalign{w_1  ~=~  & \coeff{1}{2}\,(i w_4) \, \big[ (\zeta_1^2 + \zeta_2^2)
~+~ e^{2\,h}\, (\bar \zeta_1^2 + \bar \zeta_2^2)  \big] \,, \cr 
w_2  ~=~  & -\coeff{i}{2}\, (i w_4) \, \big[ (\zeta_1^2 - \zeta_2^2)
~-~ e^{2\,h}\, (\bar \zeta_1^2 - \bar \zeta_2^2)  \big] \,,\cr 
w_3 ~=~ &- \coeff{i}{2}\,(i w_4) \, (\zeta_1\, \zeta_2  ~+~ e^{2\,h}\,  \bar \zeta_1\, 
\bar \zeta_2 )  \,,  \qquad w_4 ~=~ - i u\,e^{i \phi}\,.}}

One of the important lessons in this coordinate change exercise is that the master
function, $h$, appears as a natural coordinate on the conifold, and
one may view our approach to finding the Ricci-flat metric as prescribing
the metric functions and then solving equations to determine
the complex coordinates.   The simplification
that the ``master equation'' provides over the Monge-Amp\`ere equation may thus
be viewed as a clever re-definition of dependent and independent variables.

\subsec{The conifold again}

It is relatively easy to obtain the conifold solution from our re-formulation
of the problem of finding Ricci-flat metrics:  One simply sets \metAnsatz\ equal
to \rndmet\ and extracts the necessary re-definitions.  The easiest equations follow
from angular parts of the metric and one immediately finds that:
\eqn\firstconeqn{ h ~=~ \log\Big[\tan\big({\theta \over 2} \big)\Big]\,, }
along with 
\eqn\secondconeqn{ v^2~=~  {2 \over 3\, c_0}\, \rho^2 \,\cos \theta \,,}  
and $c_1=0$. 
Given this, it is then easy to find the other change of variable:
\eqn\seconduueqn{ u^4~=~   {4 \over 81\, c_2\, c_0^2 }\, \rho^6 \,\sin^4 \theta \,,}  
and show that the metrics match perfectly for any value of $c_0$ and $c_2$ if and only if $\alpha = 2$ in \mfunction.   
Thus the master function \firstconeqn\ satisfies:
\eqn\specmastereqn{ {1 \over  u^{3} } \, \partial_u \big(u\, \partial_u h \big) ~-~
{c_2 \over c_0 \,v} \,\partial_v \bigg({1\over v^3 }\,
\sinh(2\,h) \, \partial_v h \bigg) ~=~ 0\,.}
Since the dependence on $c_0$ and $c_2$ in \secondconeqn-\specmastereqn\ can be removed by rescaling $u$ and $v$, we can simply set $c_0=c_2=1$.

One can also derive this result {\it ab initio}.   That is, one can
seek a metric that is a cone over a five-dimensional Einstein space.
that depends upon a single variable $\theta$. Denote  the radial coordinate on 
the cone by $\rho$, then by taking ratios of metric coefficients one
sees  from \metAnsatz\ that $h$ can only be a function of $\theta$, and
so $v  =  \rho\, q_2(\theta)$ for some function $q_2$ provided $c_1=0$.   It thus follows
that $H_5$ and hence $H_2$ will be independent of $\rho$.  Then,
from the $d \phi^2$ terms we see that $H_1 u = \rho q_3(\theta)$,
for some function, $q_3$.  Using this in the   expression for 
$H_1^2 /H_2^2$ that one obtains from \thirdset\ one finds:
\eqn\uvbehave{u ~=~   \rho^{6 \over \alpha+2} \, q_1(\theta) \,, \qquad
v  ~=~  \rho\, q_2(\theta)\,.}
In particular, one finds that 
\eqn\xvardefn{x ~\equiv~    {u^{{1 \over 2}\,(\alpha+2)} \over 
v^3} \,,}
is independent of $\rho$, and is only a function of $\theta$.  One
can thus take $h = h(x)$, and substitute into the master equation.
For $\alpha =2$ and  $c_0=c_2=1$, this yields:
\eqn\redmaster{  \partial_x \big(x\, \partial_x h \big) ~=~ \coeff{3}{8}\, \,x^2 \,
\big( 7\, \partial_x (\cosh(2\,h))  + 3 x\, \partial_x^2 (\cosh(2\,h))   \big)    \,.}

\subsec{Monge-Amp\`ere -- Tian-Yau}

The master equation \mastereqn\ is a quasilinear PDE, but it is  very hard to 
solve analytically. However, Tian and Yau 
\refs{\tianyaua,\tianyaub} proved some existence and uniqueness theorems for 
non-compact Calabi-Yau metrics. Here we will remark upon how one 
might  employ their methods to prove that
there is a solution to the master equation with the desired properties.

One can construct a closed K\"ahler form, $J_0$, with the right asymptotics on the 
noncompact Calabi-Yau manifold. This K\"ahler form defines the K\"ahler class
of the desired solution. Since the manifold is Calabi-Yau, there exists
a holomorphic $(3,0)$-form $\Omega$. This 3-form is not unique, since the space
is noncompact, but we can fix $\Omega$ by requiring the right asymptotics and 
symmetry properties. 

A Ricci-flat, K\"ahler metric has the property
\eqn\ricciflat{{1 \over 3!}\,J\wedge J\wedge J=\Omega\wedge\overline\Omega\,.}
The two-form, $J_0$,  typically does not satisfy this equation. However 
one can try to find  a potential $\cK$, such that
\eqn\ricciflatb{{1 \over 3!}\,(J_0+\partial\bar\partial\cK)\wedge(J_0+
\partial\bar\partial\cK)\wedge
(J_0+\partial\bar\partial\cK)=\Omega\wedge\overline\Omega\,.}
This equation is a Monge-Amp\`ere equation, which is much more difficult than
our master equation. However, Tian and Yau show, that there is 
a unique potential $\cK$ with the right fall-off 
properties at infinity. The proof of this theorem respects all the symmetries
of the problem, but it does not allow for a singularity at a finite point. There 
should however be a simple generalization of the proof, where the asymptotics of 
the K\"ahler form at the singularity is specified.

We will now argue, that there is a $J_0$ with the right asymptotics at the 
singularity and at infinity. First, there is the singular conifold metric. This 
metric has the right asymptotics at the singularity. On the other hand there is
the degenerate limit of the conifold, which is $\IC^2/\ZZ_2 \times \IC$. One can now 
use a bump function to glue the two K\"ahler forms to a single K\"ahler form with
the right asymptotics. The modification of the Tian-Yau theorem should then prove 
the existence and uniqueness of a Ricci flat K\"ahler metric with the desired 
asymptotics on the singular conifold.

\subsec{Completing the  solution}

We have been implicitly using the results of \refs{\BeckerGJ,\GranaXN} 
that imply that the internal metric we seek must be a Calabi-Yau geometry.
The complete solution must involve non-trivial, harmonic warp factors and
related five-form fluxes as in \ffromhzero\ and \fiveformflux.  This, in principle,
requires us to solve the Laplace equation for the metric described above.
Amusingly, this equation, while linear, is not greatly different from the master 
equation itself.  However, without knowing the function, $h$, we do not appear
to be able to find the requisite solution analytically.   On the other hand, 
we know that we need an $O(3) \times U(1)$ symmetric, smooth harmonic 
function on this space with the appropriate asymptotics at the singularity and at 
infinity. Again, there should be  a unique such function.

All of this strongly suggests that the Klebanov-Witten flow is actually captured by 
our master equation.

\newsec{The geometry of the PW solution}

There is a striking similarity between  the analysis above and the analysis
of more general families of flows that involve fluxes 
\refs{\GowdigereJF,\PilchJG,\PilchYG,\NemeschanskyYH}. Moreover
 the master equation derived above is very similar to the equations 
 that govern these more general flows.  Given this, we now wish to revisit
 one of these solutions to elucidate the underlying geometry in the
 light of what we have seen above.  
 
 The flow we consider is, of course, the close cousin of the Klebanov-Witten
 flow in which the mass perturbation is of the form \massdef\ but with both
 terms having the same sign.  This flow lies purely within the untwisted
 sector of the gauge theory, and its holographic dual involves non-trivial
 three-form fluxes.  The $SU(2) \times U(1)$ invariant families of
 flow solutions of this type were analyzed in   \PilchYG.

\subsec{The elements of the PW solution}

We begin by recalling some of the details of the solution of
\PilchYG, but we use slightly modified conventions.  We start by recasting the
metric Ansatz in the form:
\eqn\metrans{
ds^2~=~ H_0^2 (\eta_{\mu \nu}\, dx^\mu \, dx^\nu)~-~ H_0^{-2}\, ds_6^2 \,,}
where $ds_6^2$ is given by \metAnsatz\ with $H_3 =H_4 = {1 \over 2} v$.  Note the inclusion 
of the factor of $H_0^{-2}$ in front of $ds_6^2$.  We thus use the frames:
\eqn\newframes{\eqalign{e^a ~=~ & H_0 \,  dx^a \,, \ \ \  a=1, \dots, 4 \,,  \qquad  
e^5 ~=~ H_0^{-1}\, H_1 \, du  \,, \qquad e^6 ~=~  H_0^{-1}\, H_2 \, dv \,, \cr
e^7 ~=~ & \coeff{1}{2} \, H_0^{-1}\,    v\, \sigma_1\,, \qquad e^8 ~=~ 
 \coeff{1}{2} \, H_0^{-1}\,   v\, \sigma_2\,, \cr
e^{9}  ~=~ &  \coeff{1}{2} \, H_0^{-1}\,  H_5 \, v\,  ( \sigma_3+H_6\,d\phi)
\,,  \quad e^{10} ~=~  H_0^{-1}\,  H_1\,  u\, d\phi \,,}}
Following  \PilchYG, we define: 
\eqn\Psidefn{ \Psi ~=~ \log\left(v^2\,{H_1\over H_2}\right)\,. }
Then the metric coefficients of $ds_6^2$ are determined by: 
\eqn\firstsolelts{H_1^2 ~=~  {1 \over2\, v^3} \,  e^{2 \, \Psi}\, \partial_v \Psi \,,
\qquad  H_2^2 ~=~ \coeff{1}{2} \, v\,\partial_v \Psi \,,
\qquad H_5 ~=~  H_2^{-1}\,, \qquad  H_6  ~=~  u\,\partial_u \Psi \,,
\qquad  \,.}

\subsec{The complex geometry of the PW solution}

Observe  that, with the exception of the expression
for $H_1$, the PW solution is identical to the form of the metric discovered
in sections 4 and 5.  Indeed, if one takes:
\eqn\hmatchPsi{ h~=~ \Psi ~+~ k \,,} 
and takes the limit $k \to \infty$ while scaling $c_2$ appropriately,
then the metric of section 4 matches the form of the metric
above.   

Thus, the metric in \PilchYG\ is hermitian with  an {\it integrable}
complex structure, with (1,0)-forms
\eqn\PWholforms{\omega_1  ~=~   du -  i u\, d\phi \,, \qquad
 \omega_2    ~=~  \,\sigma_1 ~+~ i\,  \sigma_2\,, \qquad
\omega_3   ~=~  d\Psi ~+~  i\, \sigma_3 \,.}
The corresponding holomorphic coordinates are:
 \eqn\zcoords{z_1 ~=~ u\, e^{- i\, \phi} \,, \quad  z_2 ~=~ e^{{1\over 2}\,
 \Psi } \,  \cos(\coeff{1}{2} \varphi_1)  \,e^{{i \over 2}\,
 (\varphi_2 + \varphi_3)} \,, \quad z_3 ~=~ e^{{1\over 2}\,
 \Psi } \, \sin(\coeff{1}{2} \varphi_1)  \,  e^{-{i \over 2}\,  (\varphi_2 - \varphi_3)} \,.}
The putative almost complex structure:
\eqn\newacstr{J ~=~ - H_1^2 \, u \, du \wedge d\phi ~+~   \coeff{1}{4}\,  v^2 \, 
\sigma_1 \wedge \sigma_2 ~+~   \coeff{1}{2}\,    v  \,  dv \wedge 
(\sigma_3 + u\,\partial_u \Psi) }
is thus a true complex structure, and the underlying complex manifold is 
simply $\IC^2/\ZZ_2 \times\IC$ parametrized by \zcoords.

Moreover, the $(3,0)$-form:
\eqn\threezeroform{\Omega ~=~  e^{\Psi -i\,\phi} \, \omega_1 \wedge \omega_2 
 \wedge \omega_3  ~=~ - \coeff{1}{4}\, d z_1 \wedge  d z_2 \wedge  d z_3 \,, }
is manifestly closed, and thus holomorphic.  It also satisfies the Calabi-Yau
condition:
\eqn\CYcond{\Omega \wedge \overline \Omega  ~=~  
\coeff{64}{3}\,i\,J \wedge J \wedge J \,,}
and thus there is a holomorphic $(3,0)$ form that is a square root of 
the volume form.  We thus have a Calabi-Yau manifold, {\it
except} that the manifold is {\it not} K\"ahler.  The complex structure (or, more
correctly, the K\"ahler form) is not closed.  

The solution of \PilchYG\ requires that $\Psi$ satisfy:
\eqn\PWmastereqn{
u^3 {\partial\over\partial u}\left({1\over u^3}\,
{\partial\over\partial u}\,\Psi\right) +
{1\over v}\,
 {\partial\over\partial v}\left({1\over v^3}\,
e^{2\Psi} \, {\partial \Psi \over\partial v}  \right)~=~ 0 \,,
}
whereas the K\"ahler condition, or Ricci-flatness of the metric,
would require
\eqn\CYmastereqn{
{1\over u} {\partial\over\partial u}\left(u \,
{\partial\over\partial u}\,\Psi\right)+ 
{1\over v}\,
 {\partial\over\partial v}\left({1\over v^3}\,
e^{2\Psi} \, {\partial \Psi \over\partial v}  \right)~=~0\,.
}

Thus the solution of  \PilchYG\ is based upon a the
complex manifold $\IC^3$, endowed with hermitian, non-K\"ahler
metric.   The mysterious master equation of \PilchYG\ thus turns out to be
simple deformation of the K\"ahler condition.  

It is also worth noting that if one sets $h =\pm  (\Psi + k)$, where
$k$ is a constant,  in  \mastereqn\ 
and takes the limit $k \to \infty$ with $c_2 \to 0$ then \mastereqn\
(for $\alpha=2$ and $c_1 =0$) degenerates to \CYmastereqn.
This substitution of $h = \pm (\Psi + k)$ and limit of $k$ is precisely the
limit that degenerates the conifold to the orbifold, and so it is to be
expected that the ``master equation'' degenerate to the Ricci-flat
condition for $\IC^2/\ZZ_2 \times \IC$, and hence equation
\CYmastereqn.

\newsec{Conclusions}

Our purpose has been to elucidate the geometry that underlies the
holographic duals of $\cN=1$ supersymmetric flows from $\cN=2$ quiver gauge 
theories.  We are particularly interested in flows generated by mass terms
for the chiral matter multiplets in the $\cN=2$ vector multiplets.   The 
twisted sector masses are dual to blow-up modes in the dual geometry
while the `total' mass parameter comes from the untwisted sector
and is dual to a three-from flux.   Here, for simplicity, we
have focussed on the $\widehat A_1$ quiver theory, but we believe 
that our results apply in general.  Moreover, we studied the two 
extreme cases: a flow purely in the twisted sector (the Klebanov-Witten flow)
and the PW flow.

We have argued that flows purely in the twisted sector must have an underlying
Calabi-Yau geometry {\it all along the flow}, and that the complete solution is
simply an extension via the harmonic Ansatz.  For the $\widehat A_1$ quiver
we showed that this means that the underlying manifold must be the 
singular conifold with a Calabi-Yau metric that interpolates between 
the orbifold point at infinity and the cone over $T^{(1,1)}$ in the deep interior.
The standard approach to finding such a metric would involve the 
solving the Monge-Amp\`ere equation, a sixth order, non-linear
PDE.  However, by using the symmetries and structure of the flow we
were able to make an Ansatz of sufficient generality to capture the solution
and yet reduce everything to the solution of a single, second order
quasi-linear PDE.   This substantial simplification compared to the 
Monge-Amp\`ere approach arises partially because we have a higher
level of symmetry, and do not need the full force of the general method.
The other reason for the simplification is because of an inversion of
dependent and independent variables:  Monge-Amp\`ere approach
leads to differential equations for metric functions in terms of fixed
complex coordinates, while the ``master equation'' is really a differential
equation for part of the complex coordinates having fixed some of the metric
functions.  At any rate, the ``master equation'' is, on the face of it,  far simpler 
than the Monge-Amp\`ere equation, and the master equation has a very
simple linearization that leads to an obvious perturbation expansion.

One of the other consequences of this description of a Calabi-Yau
manifold by the master equation \CYmastereqn\ is that it sheds 
some light upon other non-Calabi-Yau, flux solutions, like those of
\refs{\GowdigereJF,\PilchJG,\PilchYG,\NemeschanskyYH}, that are
completely determined by a very, very similar such equation.
This strongly suggest that such solutions must be relatively simple
modifications of the Calabi-Yau condition.   To understand this better, we 
re-examined the geometry underlying the solution of \PilchYG,  and showed that
this solution is essentially a a non-K\"ahler Calabi-Yau manifold:
It has an integrable complex structure, a hermitian
metric, and a holomorphic $(3,0)$ form that is the square root of the
volume form.  We are continuing the study of this geometry and its
generalizations, and we would particularly like to understand the
geometric meaning of the simple deformations of the master equations
that take one from the Calabi-Yau solutions to the non-trivial
flux solutions.

It is also useful to recall that the master equation, \PWmastereqn\ 
possesses a known, highly non-trivial solution coming from gauged 
supergravity \FreedmanGP.  That is, there is a steepest descent flow
on a simple five-dimensional superpotential  that provides one of
the most interesting flows to the non-trivial IR fixed point,
and while this solution is not given analytically, its solution is
trivially obtained graphically.   If one could forge the connection 
between the various master equations more deeply, one might hope
to find special Calabi-Yau solutions by such simple, graphical techniques.

There is also the work of \refs{ \CorradoWX, \CorradoBZ} 
concerning the continuous family of flows involving arbitrary mixtures
of mass terms in the twisted and untwisted sectors.  Moreover, it was
argued in  \CorradoWX\ that all these flows should have a continuous
family if IR fixed points.   We are presently generalizing the results presented
here so as to try to capture this broader family of flows.  

We hope, ultimately, to be able to give a complete geometric characterization
of supersymmetric backgrounds involving fluxes.  It is evident from the
results presented here and elsewhere that trying to characterize the
supersymmetry bundle directly is a very effective approach in studying
not only these problems, but also in re-visiting Calabi-Yau geometries.

\bigskip
\leftline{\bf Acknowledgements}

This work was supported in part by funds
provided by the U.S.\ Department of Energy under grant number 
DE-FG03-84ER-40168. In addition, N.H. has been supported by the Fletcher-Jones graduate fellowship at USC, while K.P. has been supported in part by the grant \# SAB2002-0001 from the 
Ministerio de Educaci\'on, Cultura y Deporte of Spain.  K.P. would like to thank  the members of the Departament ECM  for their hospitality during his 
sabbatical at the University of Barcelona. We would also like to thank R.~Corrado
for helpful conversations.
 
 \vfill\eject
 \listrefs
  \vfill\eject
 \end